\documentclass[aps,prc]{revtex4}
\usepackage{amssymb,amsmath,epsfig, psfrag, ulem, color} 

\newcommand{\ee}{\end{equation}}
\newcommand{\be}{\begin{equation}}
\newcommand{\bea}{\begin{eqnarray}}
\newcommand{\eea}{\end{eqnarray}}
\newcommand{\bean}{\begin{eqnarray*}}
\newcommand{\eean}{\end{eqnarray*}}

\newcommand{\gapproxeq}{\lower
.7ex\hbox{$\;\stackrel{\textstyle >}{\sim}\;$}}
\newcommand{\lapproxeq}{\lower
.7ex\hbox{$\;\stackrel{\textstyle <}{\sim}\;$}}

\newcommand{\qq}{$q\bar{q}$~}

\newcommand{\bk}[2]{\langle\phantom{|}{#1}\phantom{|}|\phantom{|}{#2}\phantom{|}\rangle}
\newcommand{\ket}[1]{|           {#1}           \rangle}
\newcommand{\bra}[1]{\langle           {#1}           |}       
\newcommand{\mel}[3]{\bra{#1}{#2}\ket{#3}}
\newcommand{\ts  }{^3\textrm{S}_1}
\newcommand{\tp  }{^3\textrm{P}_0}
\newcommand{\vc}{\mathbf}

\def\3bar{$\bar {\hbox{\bf 3}}$}
\newcommand{\cleb}[6]{\langle{#1}\;{#2},{#3}\;{#4}|{#5}\;{#6}\rangle}
\begin{document}

\title{\bf Hybrid meson properties in Lattice QCD and Flux Tube Models}

\author{
T. J. Burns\footnote{e-mail: burns@thphys.ox.ac.uk}
and F. E. Close\footnote{e-mail: F.Close1@physics.ox.ac.uk}
}

\affiliation
{ Rudolf Peierls Centre for Theoretical Physics,
University of Oxford, \\
Keble Rd., Oxford, OX1 3NP, United Kingdom}

\date{\today}

\begin{abstract}
Flux tube model predictions for hybrid meson decays are beginning to be confronted by Lattice QCD. 
We compare the two approaches for the $S$-wave decay of the exotic $1^{-+}$, and find excellent
agreement. Results suggest that in Strong QCD \qq creation occurs with $S=1$.
\end{abstract}

\maketitle


\vskip 1.cm

While Lattice QCD is now a mature guide for the masses of glueballs and hybrids, at least in the 
quenched approximation
\cite{peardon,michael}, it is not yet able to determine hadronic decays extensively.
Flux tube models of both spectra \cite{ip,bcs} and decays\cite{ikp,cp95} have been developed, in 
part stimulated by
attempts to model the lattice, and lattice work has confirmed their 
spectroscopy\cite{michael,bcs}. 
The first study of the hybrid meson
decays $1^{-+} \to \pi b_1$ and $\pi f_1$ has recently been made in lattice
QCD\cite{mm06} 
and shows features that had been anticipated in flux-tube models\cite{ikp,cp95}. 
In this paper we compare these results and assess the implications.

Lattice QCD enables the properties of strong QCD to be explored
``experimentally" over a kinematic and parameter space that is richer than the ``physical"
values to which
nature has restricted us. Thus, for example, ref \cite{mm06} is restricted to $S$-wave decays, 
which 
it achieves with specially chosen 
masses so as to produce a decay at rest. To convert the results
to widths 
(presented in Table \ref{ftvlat}) a large extrapolation was made from the threshold, relying on
an assumption that is beyond lattice QCD, thereby masking its primary results.
It is tantalising that the ratio of widths to $\pi b_1$ and $\pi f_1$ in lattice QCD and the 
flux tube
simulations agree whereas the absolute widths, as presented in ref\cite{mm06}, are rather larger 
than those of refs\cite{ikp,cp95}. The lattice QCD
results for the known decay $b_1 \to \omega \pi$ also exceed data
when the extrapolations of ref \cite{mm06} are employed. 

This {\it prima facie} suggests that the spin-dependent features of Strong QCD
as revealed by the lattice are contained within the flux tube model, but that the momentum
dependence of the assumed extrapolation differs.  
We show here that this is the case; that when the flux tube model is
applied in the $k \to 0$ limit of the lattice
the agreement is excellent; and that the results of lattice QCD reinforce the flux-tube hypothesis 
that
\qq creation is spin triplet.

\section*{Decay amplitudes in Lattice QCD and Flux Tube Models as $k \to 0$}

The flux tube model has successfully described transitions among conventional
mesons, $\cal{M}\to \cal{M} + \cal{M}$\cite{ki}, 
and also been applied to the decays of hybrid mesons, $\cal{H} \to \cal{M} +
\cal{M}$\cite{ikp,cp95,pss}.
 A notable feature of the latter, which also emerges in some other 
models\cite{pene},
is that the prominent decays are to excited mesons, notably $S+P$ states\cite{ikp}.
In particular the exotic $\pi_1$ $1^{-+}$ (called $\hat\rho$ in \cite{mm06}) is expected to have 
prominent decays into $\pi b_1$ 
and $\pi f_1$, with the former favoured by about a factor of four in both $\tp$ and $\ts$ 
flux-tube models\cite{ikp,cp95,pss}. For a decay of width $\Gamma$ at momentum $k$, the coupling 
constant for the process $A\to B +C$ can be expressed
\be
\Gamma/k=\frac{1}{\pi}(\mathbf{PS})\mel{BC}{\vc\sigma\cdot\vc\nabla}{A}^2 
\ee
where $\mathbf{PS}$ denotes phase space and $\mel{BC}{\vc\sigma\cdot\vc\nabla}{A}$ is the overlap 
of the quark and string wavefunctions coupled by the $\tp$ string breaking operator. The overlap 
is explicitly momentum dependent, and assumes the same form in each of the cases of present 
interest. Defining
\be
 S(k)=\frac{\gamma_0\pi^{3/4}2^{5/2}}{\beta^{1/2}3}\left(1-\frac{2k^2}{9\beta^2}\right)e^{-k^2/12\beta^2}
\ee
the relevant $S$-wave amplitudes can be written \cite{ikp,cp95}
\bea
\mel{\omega\pi}{\vc\sigma\cdot\vc\nabla}{b_1}&=& S(k)\label{b1decay}\\
\mel{b_1\pi}{\vc\sigma\cdot\vc\nabla}{\pi_1}&=&\sqrt{2}\frac{\kappa\sqrt{b}}{\beta} 
S(k)\label{pi1decay1}\\
\mel{f_1\pi}{\vc\sigma\cdot\vc\nabla}{\pi_1}&=&\frac{1}{\sqrt 2}\frac{\kappa\sqrt{b}}{\beta} 
S(k)\label{pi1decay2}
\eea
 for harmonic oscillator wavefunctions of equal width $\beta$ and where $\gamma_0$ is an overall 
pair creation constant fit to data. We defer 
 further discussion of the parameters temporarily, noting only that the relative scale of hybrid 
decays to that of conventional mesons is driven by the string tension $b$ and a factor $\kappa$ 
that emerges from the overlap of the excited flux tube with ground state flux tubes: thus in the 
flux tube model there is an immediate correlation of scale between the decay widths of 
conventional and hybrid states. 

\begin{figure}
\begin{center}
\includegraphics[angle=-90,width=0.6\textwidth]{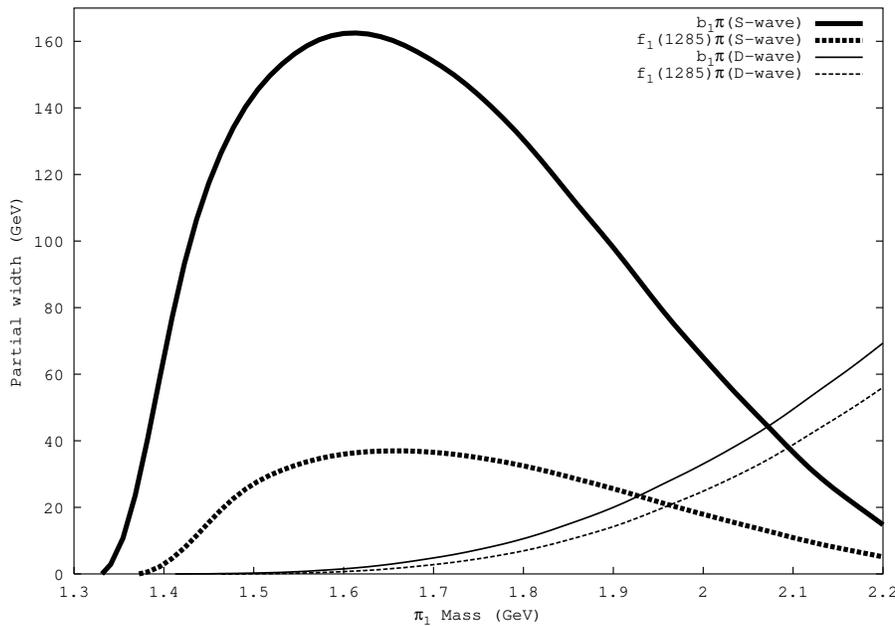}	
\end{center}	
\caption{Partial widths of  $\pi_1$ as a function of mass in the flux tube model with parameter 
set (A).}
\label{pi1}
\end{figure}

In Table \ref{ftvlat} we present flux tube predictions for the $\pi_1$ width and compare
with results extrapolated from lattice QCD in ref\cite{mm06}. 
Ref \cite{ikp} 
calculated the dominant decay amplitudes of exotic-$J^{PC}$ hybrid states with a localised pair 
creation region, quoted results for a $\pi_1$ state 
at 1.9GeV. 
Ref \cite{cp95} verified those results in an infinite flux tube approximation and calculated 
analytic forms for both exotic 
and non exotic hybrid decays as a function of mass. We quote the results for a $\pi_1$ at 2.0 GeV 
 having corrected
some numerical factors in ref \cite{cp95}, and present the amplitudes for arbitrary $\pi_1$ mass in Fig. \ref{pi1}. The 
model
predicts widths for the $1^{-+}$ decays in both $S$ and $D$ waves and as a function of the hybrid 
mass, 
with 
the $S$-wave contributions dominating. 

\begin{table}

\begin{tabular}{lccc}
\hline
				&FTM		&FTM		&Lattice		\\
				&Ref \cite{ikp}	&Ref\cite{cp95}	& Ref \cite{mm06}	\\
				& 1.9GeV	& 2.0 GeV	& 2.0 GeV	\\
\hline
$\Gamma(\pi_1\to b_1\pi)_S$	&100		&70		& $400\pm 120$		\\
$\Gamma(\pi_1\to b_1\pi)_D$	&30		&30		&		\\
$\Gamma(\pi_1\to f_1\pi)_S$	&30		&20		& $90\pm 60$  	\\
$\Gamma(\pi_1\to f_1\pi)_D$	&20 		&25		&		
	\\
\hline
\end{tabular}
\caption{Comparison of flux tube and lattice predictions for $\pi_1$ decays. }
\label{ftvlat}
\end{table}

 The lattice technique is to put 
a given decay channel at roughly the same energy as the decaying state so that the decay is just 
allowed while conserving energy in a two-point function \cite{mm06}. For a lattice with spacing 
$a$ and size $L$, the lattice transition amplitude $xa$ gives the analogue of equations 
(\ref{b1decay})--(\ref{pi1decay2}):
\be
\mel{BC}{\textrm {Latt}}{A}=(L/a)^{3/2}(xa)\label{lattcoupling}\\
\ee
so that the coupling constant at threshold, with phase space {\bf PS} chosen appropriately with 
lattice masses and a decay at threshold, can be written
\be
\Gamma/k=\frac{1}{\pi}(\mathbf{PS})\mel{BC}{\textrm{Latt}}{A}^2.
\ee

In order to make a statement about physical widths, ref.\cite{mm06} assumes that $\Gamma/k$ 
doesn't vary with quark 
mass.   This linear extrapolation leads to the large width of $\Gamma=400\pm 120$ MeV in Table 
\ref{ftvlat} for a $\pi_1$ at 2.0GeV decaying to $b_1\pi$ with physical masses. 
Ref \cite{mm06} also presented results for the conventional decay $b_1\to \omega\pi$ noting that 
an equivalent extrapolation overestimates the data and that this could be generic. We note the 
result would be $\sim 220$MeV, significantly larger than the $S$-wave data $\sim 130$MeV 
\cite{pdg}. 


On the other hand, the flux tube model extrapolation has been tested over a large range of $k$, 
predicting accurately the decays of both mesons and baryons \cite{ki,bcps,cs05}. 
For the physical $b_1$ at 1235MeV decaying to $\omega\pi$ the model reproduces very nicely the experimental data for both $S$-wave $\sim 130$MeV (using equation (\ref{b1decay})) and $D$-wave $\sim 10$MeV. The successful phenomenology of this and a wide 
range of other conventional meson decays  relies on momentum-dependent form factors arising from 
the overlap of hadron 
wavefunctions. The need for such form factors is rather general,
empirically supported as exclusive hadron decay widths do not show unrestricted
growth with phase space\cite{pdg}. Such phenomena are also expected for hybrid decays
$\cal{H} \to \cal{M} +\cal{M}$ and appear explicitly in equations (\ref{pi1decay1}) and 
(\ref{pi1decay2}).  

The extrapolation from the 
lattice limit $k \to 0$ assumed in ref \cite{mm06}
ignores any such $k$ dependent suppression: thus the predicted widths are much larger and
for $b_1 \to \omega \pi$ disagree with experiment. 
Such an assumption may apply for inclusive decays but is unphysical for exclusive channels as 
here.
As the momentum $k$ increases, individual channels fall at the expense of multi-body channels 
opening (this is the
physics of exclusive form factors) even though the sum of channels may be $k$ independent (scale 
invariant)
\cite{ci02}. More generally, the hadron size sets an explicit scale against which the momentum
$k$ of the exclusive process is weighed; a linear extrapolation ignores this.

 Another difference is the functional parametrisation of phase space
$(\mathbf{PS})$  in refs \cite{ikp,cp95} and \cite{mm06}, respectively
\be
(\mathbf{PS})_{ft}=\frac{\widetilde{M}_B\widetilde{M}_C}{\widetilde{M}_H},
\qquad\qquad
(\mathbf{PS})_{lat}=\frac{aE_BaE_C}{aE_B+aE_C}
\ee 
where $\widetilde{M}$ are meson masses calculated before spin interactions (for a discussion 
see Appendix A2 of ref \cite{cs05}) and $aE$ are the lattice masses. The latter phase space is for 
a decay at threshold: since the physical decay is far from threshold, 
$(\mathbf{PS})_{ft}<(\mathbf{PS})_{lat}$. 

Thus there is no direct comparison between the widths of
the lattice predictions in  Table \ref{ftvlat} 
and those of the flux tube model: the former works with unphysical masses and 
calculates the amplitude at threshold, extrapolating  to the large momenta required for 
the physical masses; the latter calculates the amplitude with physical masses far from threshold,
 with dynamics at this momenta determined by the overlap of quark and string wavefunctions.

To compare the two approaches we evaluate the flux tube predictions for $k=0$ and compare not the 
coupling constants $\Gamma/k$ but the transition amplitudes $\mel{BC}{\vc\sigma\cdot\vc\nabla}{A}$ 
and $\mel{BC}{\textrm{latt}}{A}$. The lattice couplings follow from the slopes $(xa)$ recorded in 
ref \cite{mm06}; we list these couplings, for the two different codes, in the first two rows of 
Table \ref{gammas}.   The analagous flux tube couplings follow from equations 
(\ref{b1decay})--(\ref{pi1decay2}) without further assumption, and these 
are shown in the remaining rows of Table \ref{gammas} for two ``standard'' parameter sets 
(discussed below).
The overall scale of decays in the flux tube model is driven by the pair creation 
constant $\gamma_0$ and the hadronic wavefunction width $\beta$, both of which are strongly 
constrained by data. The model then fixes the scale of decays involving hybrid mesons according to 
the overlap of the string degrees of freedom, exhibited in the ratio $\kappa\sqrt b/\beta$.

\begin{table}
\begin{tabular}{l rl rl rl}
\hline
				
			&$b_1$	&$\to\omega\pi$ &$\pi_1$&$\to b_1\pi$	 &$\pi_1$	&$\to f_1\pi$\\
\hline
Lattice	 (C410)		&2.3 	&$\pm$0.1	&2.9 	&$\pm 0.4	$&1.5 		&$\pm0.4$		\\	
Lattice	 (U355)		&3.4 	&$\pm$0.2	&2.9 	&$\pm 0.3	$&1.1 		&$\pm0.4$		\\
\hline
Flux tube (A)		&2.7   &        	&2.9    &           	&1.4 &        		\\
Flux tube (B) 		&3.3   &         	&3.9    &            	&1.9 &         		\\
\hline
\end{tabular}
\caption{Transition amplitudes $\mel{BC}{\vc\sigma\cdot\vc\nabla}{A}$ and 
$\mel{BC}{\textrm{latt}}{A}$ in units of GeV$^{-1/2}$. The parameter sets (A) and (B) are\\
(A) $\gamma_0=0.39,\beta=0.40,\kappa=0.7$\\
(B) $\gamma_0=0.45,\beta=0.36,\kappa=0.7$
}
\label{gammas}
\end{table}

Before comparing the two approaches, we briefly discuss the nature of the parameter selection 
appropriate for the flux tube. The analytic expressions (\ref{b1decay})--(\ref{pi1decay2}) are 
those appropriate to a radial hybrid wavefunction $\sim r^\delta e^{-\beta^2r^2/2}$ with 
$\delta=1$; 
as noted in \cite{cp95} the results differ very little from the ``true'' radial wavefunction 
which, 
ignoring a term that raises/lowers the gluonic angular momentum, has $\delta\approx 0.6$. As they 
are 
presented, the expressions (\ref{b1decay})--(\ref{pi1decay2}) correspond to setting 
$e^{-fby_{\perp}^2/2}=1$
 in the flux tube overlap term. The original flux tube formulation established that the inclusion 
of a 
 localised pair creation region has very little effect on the predictions, on account of the 
asymptotic 
 forms of the hadron wavefunctions automatically imposing a ``flux tube''-like structure on the 
 overlaps \cite{ki}. The approach of \cite{cp95} uses an infinitely long flux tube, but for the 
 case of equal wavefunction widths the flux tube information integrates out and merely rescales 
 the parameter $\gamma_0$. Thus in a sense the infinite flux tube approximation is equivalent to 
 the absence of a $e^{-fby_\perp^2/2}$ flux tube term altogether: the 
 naive $\tp$ model with equal probability pair creation everywhere in space has effectively been 
recovered,
  with
 a re-scaling of the overall pair creation parameter. Thus the appropriate choice of 
 the $\gamma_0$ that appears in (\ref{b1decay})--(\ref{pi1decay2}) is that which has 
 been fit to the decays of mesons in a model with no flux tube supression away from 
 the \qq axis: several such fits are available and constrain both $\gamma_0$ and $\beta$ 
 rather strongly. The authors of ref \cite{ikp} advocate $\gamma_0=0.39$ and $\beta=0.4$; 
 ref \cite{bcps} fit to a wide range of higher quarkonia and settle on $\gamma_0=0.4-0.5$ 
 and $\beta=0.36$. As is evident Table \ref{gammas}, these parameter sets 
correctly set the scale of the decay $b_1\to\omega\pi$. The spread of values in the lattice 
calculation is reproduced by the spread of ``phenomenologically-allowed'' flux tube parameters.

The relative scale of hybrid decays is then set by the ratio $\kappa\sqrt b/\beta$. 
The string tension $b=0.18$GeV$^{2}$ is rather tightly constrained, 
thus the results hinge only on the factor $\kappa$ that emerges directly from the overlap of the 
stringlike degrees of freedom of the hybrid and conventional mesons. Explicit calculation in the 
framework of the harmonic approximation \cite{dpp} has $\kappa$ as a function that depends on the 
longitudinal distance along the original \qq axis at which the string breaks, varying from $0.8- 
0.9$ 
at its peak in the centre of the meson to $0.4-0.6$ at either end, depending on the degree of 
quantization.
 Refs \cite{ikp} and \cite{cp95} treated $\kappa$ as a constant, choosing $\kappa=1$ and $ 0.9$ 
 respectively. The aforementioned discussion would suggest $\kappa\sim 0.7$ may be a more 
realistic
  implementation of the model, and indeed such a choice reproduces the lattice calculations rather 
  nicely. 

 The hadrons of the lattice are generally heavier than their experimental values, and as such it 
is not automatic that the $\beta$ fit to the decays of experimental states should reproduce the 
decays on the lattice. However, it is only the pion that is drastically different from its true 
mass; moreover, this latttice-pion is no longer exceptional and in consequence may be described by 
the same scale as other hadrons.

Finally we consider the possiblity of allowing the initial and final state wavefunction to have 
different widths. Ref \cite{cp95} prefer a slightly smaller $\beta$ for the initial hybrid state:
 such a choice manifests itself in a rescaling by $\beta_A/\beta_B$. The effect on the numbers in
  Table \ref{gammas} is not drastic, though it improves the agreement somewhat,
   particularly for parameter set (B) with its larger $\gamma_0$. 

With the state of the lattice uncertainties at present, it is not appropriate to attempt a best 
fit
 for flux tube parameters. We are very much encouraged, however, that the ``standard'' choice of 
 harmonic oscillator parameters reproduces the lattice results with remarkable accuracy. Even more
  encouraging is that the string overlap factor correctly sets the scale of hybrid decays relative
   to conventional meson decays: in this result we have at a very direct test of the flux tube 
dynamics.
    Agreement with the lattice is non-trivial: the relative
 strengths of hybrid and conventional decay amplitudes emerge naturally from the string-like 
description 
 of the gluonic degrees of freedom and are determined by the same
  string tension that controls the conventional hadron spectrum. 
  It is notable that the choice of $\kappa$ suggested by the harmonic approximation calculations 
  reproduces very nicely this relative scale. That this choice is smaller than that traditionally 
  used suggests that the previous predictions for hybrid widths should be scaled by 
  $\approx (0.7/0.9)^2\approx 0.6$, which is encouraging from the point of view of the 
  experimental hunt for these states.
  We also note that the same relative scale emerges 
  from a continuum string picture, as can be seen from the plot of the continuum limit of 
  $\kappa$ in \cite{dpp}.

Emboldened by the good agreement, we suggest an extrapolation of the lattice results at threshold
 to the physical region. To the extent that the two approaches agree at $k=0$, the relevant 
 ``extrapolation'' is explicit in equations (\ref{pi1decay1}) and (\ref{pi1decay2}). The results 
are shown in Fig \ref{pi1} with parameter set (A) (see the caption of Table \ref{gammas}), spin-averaged masses in the phase space, and an
	 $f_1$ octet-singlet mixing angle of 50$^o$ following \cite{closekirk}. The rather more modest 
widths 	 than those of the linear lattice extrapolation offer encouragement for the experimental observation of such 
	 states. We note in passing, however, that if flux tube and lattice mass estimates are a 
guide, hybrids may couple strongly to many previously unconsidered modes such as 
$1\textrm{P}+\textrm{vector}$ and $1\textrm{D}+\textrm{pseudoscalar}$. Although such modes appears 
not to spoil the rather narrow $\pi_1$, they are important for other states such as $2^{+-}$ 
\cite{tjb}.

 The predictions for isoscalar decays to $\pi a_1$ \cite{mm06} also agree in 
relative scale with the flux tube. This is trivial as only Zweig connected diagrams are considered 
and the
different modes are related primarily by flavour factors. Likewise the transition amplitudes for 
$S$-wave $KK_1$ directly follow from the aforementioned discussion, with some adjustments due to 
relative flavour factors: we find 
$\mel{K^1P_1K}{\vc\sigma\cdot\vc\nabla}{\pi_1}=\mel{b_1\pi}{\vc\sigma\cdot\vc\nabla}{\pi_1}/\sqrt 
2$ and 
$\mel{K^3P_1K}{\vc\sigma\cdot\vc\nabla}{\pi_1}=\mel{f_1\pi}{\vc\sigma\cdot\vc\nabla}{\pi_1}$.

\subsection*{The ratio of $\pi b_1$ to $\pi f_1$}

The ratio of $\pi b_1$ to $\pi f_1$ allows a direct test of the pair creation mechanism without 
entering into the nuances of parameters. The averaged lattice results of ref \cite{mm06} suggest 
that the former is favoured by a factor of 2 in amplitude (4 in width), which is tantalising in 
that both the $\tp$ and $\ts$ decays models predict such a ratio, as can be read off from 
equations (\ref{pi1decay1}) and (\ref{pi1decay2}) in the $\tp$ case. Flavour immediately accounts 
for an enhancement of $b_1\pi$ by a factor of $\sqrt 2$ in amplitude; the remaining enhancement 
observed in the lattice is therefore due to the spin and angular dynamics that differentiate the 
$b_1$ and $f_1$ (with quark spin 0 and 1 respectively).

Fundamental to the flux tube models has been the assumption that $L$ and $S$ factorise 
and that $S$ is conserved in the decay. 
The latter assumption, coupled with spin 1 pair creation, forbids decays of the type
$
(S=0)\to (S=0)+(S=0)
$
by orthogonality of the spin wavefunctions and experiment seems to support it, for instance the 
absence
  of the decay $\pi_2(1670)\to b_1\pi$. On the other hand,  $S=0$ pair 
  creation would forbid decays of the type 
$
(S=1) \to (S=0)+(S=0)$ and $
(S=0) \to (S=1)+(S=0);
$, a possibility already excluded by data: $\rho\to\pi\pi$ and $b_1\to\omega\pi$ are	 examples 
of such modes,
both well known experimentally and on the lattice.

In flux tube and constituent gluon models, the $1^{-+}$ has $q\bar{q}$ in spin 1 and so 
would not decay to $b_1\pi$ if its decay were driven by spin 0 pair creation.  Thus the lattice  
observation $\pi_1\to b_1\pi$ supports the flux tube hypothesis that hybrid 
decays are driven by the same pair creation mechanism as that of conventionals, and that a 
hybrid can be described in terms of quark $S= 0$ or 1 coupled to $L=1$ carried explicitly by the 
gluonic flux tube.

Spin 1 pair creation is natural in the flux tube model, with both  $\ts$ and $\tp$ operators 
emerging from expansion of the strong coupling Hamiltonian. In the extremely strong coupled limit 
where the flux tube is straight, pair creation occurs along the interquark axis  in 
$\ts$ 
via the operator $\vc \sigma\cdot \hat r$. In the original formulation of the model it was argued 
that zero point oscillations of the flux tube  will wipe out this term leaving instead $\tp$ 
pair
 creation via the operator $\vc \sigma\cdot \vc \nabla$, and this was found to give better 
agreement 
 with experiment \cite{ip}. In either mechanism, there is a common feature that may well be 
verified
  by the lattice: the decay of $\pi_1 \to \pi b_1$ is dominant over $\pi f_1$ by a factor of four, 
  independently of momentum \cite{ikp,cp95,pss,mm06}. The origin of the dynamical effect is 
  not readily explained by the lattice calculation; in the flux-tube model, as we now demonstrate, 
it emerges naturally from  spin 1 pair creation by a scalar operator, and subsequent recoupling to 
final 
  state mesons. 

As shown in equations (4) and (5) of \cite{cp95}, the flux tube model decay amplitude can be 
written as 
a linear combination of spatial overlaps $I_{M_L^AM_L^B}$, that combination determined by angular 
momentum 
and partial wave recoupling of the initial and final states. The $I_{M_L^AM_L^B}$ are the matrix 
elements 
of the  pair creation operator between the initial and final quark and string wavefunctions. 
Including 
flavour and spin wavefunction overlaps as in eqn. (3) of \cite{cp95}  gives
\bea
S(k)=\sqrt 3 \bk{\phi_B\phi_C}{\phi_A\phi_0}
\sum_{M_L^AM_S^A\lambda}\cleb{1}{M_L^A}{1}{M_S^A}{1}{M_L^A+M_S^A}\cleb{1}{\lambda}{1}{-\lambda}{
0}{0}\\
\cleb{1}{M_L^A+\lambda}{S_B}{M_S^A-\lambda}{1}{M_S^A+M_L^A}
\bk{\chi_{S_B}^{M_S^A-\lambda}\chi_0^0}{\chi_1^{M_S^A}\chi_1^{-\lambda}}I_{M_L^A,M_L^A+\lambda}(k)
\eea
where  $\bk{\phi_B\phi_C}{\phi_A\phi_0}$ and 
$\bk{\chi_{S_B}^{M_S^A-\lambda}\chi_0^0}{\chi_1^{M_S^A}\chi_1^{-\lambda}}$ are the overlaps of the 
flavour and spin wavefunctions of the initial meson ($q_1\overline{q_2}$) and the emergent pair  
($q_3\overline{q_4}$) reordered to give final states ($q_1\overline{q_4}$) and 
($q_3\overline{q_2}$), and the $q_3\overline{q_4}$ emerge in either $\ts$ or $\tp$ with 
$\ket{S,S_z}=\ket{1,-\lambda}$. With  $\phi_0=(u\overline u+d\overline d+s\overline s)/\sqrt 3$ 
the flavour overlap is $1/\sqrt 3$ for $b_1\pi$ and $1/\sqrt 6$ for $f_1(n\overline n)\pi$. 
The remaining algebra is tedious but straightforward giving, apart from common numerical factors,
\bea
\mel{b_1\pi}{\vc\sigma\cdot\vc\nabla}{\pi_1}&=&+2\sqrt{1/6}(I_{++}-I_{0+}+I_{+0})\\
\mel{f_1\pi}{\vc\sigma\cdot\vc\nabla}{\pi_1}&=&-\sqrt{1/6}(I_{++}-I_{0+}+I_{+0})
\eea
Crucial to the above result is the spin wavefunction of the emergent \qq pair. For the $f_1\pi$ 
mode, the Clebsch-Gordan factor $\cleb{1}{M_L^A+\lambda}{1}{M_S^A-\lambda}{1}{M_S^A+M_L^A}$ plays 
a role and brings $\pm \sqrt {1/2}$. This factor changes sign under $(M_L^A,M_S^A,\lambda)\to 
(-M_L^A,-M_S^A,-\lambda)$, matched by the changing sign of the spin overlap term 
$\bk{\chi_{1}^{M_S^A-\lambda}\chi_0^0}{\chi_1^{M_S^A}\chi_1^{-\lambda}}$ under the same operation. 
These compensating signs yield the same linear combination of $I_{++},I_{0+},I_{+0}$, scaled 
by $- \sqrt {1/2}$. The end result is that in spin triplet creation models 
\be
\frac{\Gamma(\pi_1\to b_1\pi)}{\Gamma(\pi_1\to f_1\pi)} = 4 \label{ratio}
\ee
(apart from small phase space and $k$ dependent corrections.) We stress that this result is 
independent of the spatial overlaps, and as such is independent of the detailed forms of the quark 
and string wavefunctions. Furthermore, the result is characteristic of both $\ts$ and $\tp$ 
models, being driven by the same angular momentum algebra, and depends crucially on the spin 1 
nature of the emergent \qq pair. The same linear combination drives all of the $S$-wave decays of 
hybrids with negative parity to $1\textrm{P}+^1\textrm{S}_0$ modes, and the relevant recoupling 
coefficients can be 
read from Table 1 of \cite{cp95}, combined with the appropriate flavour overlaps.

\subsection*{Conclusions}

These results show that near threshold lattice QCD and flux tube models are in excellent 
agreement. It is possible to compare quite directly the flux tube model with the underlying 
QCD, in part because one of the main uncertainties in any calculation of this type (phase 
space) has 
been removed. The standard quark model parameters give excellent agreement with the lattice 
results, 
and the encouraging agreement with the scale of hybrid decays supports the physical picture of the 
string wavefunction. For physical widths, momentum dependent form factors are crucial and we 
suggest 
that the
widths of ref \cite{mm06} are overestimates. 

Any improvement of the uncertainties in the lattice ratio of $b_1\pi:f_1\pi$, or the calculation 
of any analagous ratios, would be a welcome advance. This will allow the decay mechanism to be 
probed rather directly: more general decay models, such as those 
trigerred by the emission of a single vector gluon with the possibility of spin flips, do not 
result 
automatically in the ratio of eq. (\ref{ratio}). Finally we suggest that hybrid decays of the type 
$0^{+-},2^{+-}\to \rho\rho$ may be worth investigating on the lattice: these modes should not 
require drastic extrapolation to the physical regime, and are predicted to vanish exactly in 
several variants of quark models.

We acknowedge discussions with J Dudek. This work is supported,
in part, by grants from
the Particle Physics and
Astronomy Research Council, the Oxford University
Clarendon Fund and the
EU-TMR program ``Eurodice'', HPRN-CT-2002-00311.


\end{document}